\documentstyle[12pt,aaspp4]{article}
\begin{document}

\newcommand{\up}[1]{\ifmmode^{\rm #1}\else$^{\rm #1}$\fi}
\newcommand{\zdot}{\makebox[0pt][l]{.}}
\newcommand{\upd}{\up{d}}
\newcommand{\uph}{\up{h}}
\newcommand{\upm}{\up{m}}
\newcommand{\ups}{\up{s}}
\newcommand{\arcd}{\ifmmode^{\circ}\else$^{\circ}$\fi}
\newcommand{\arcm}{\ifmmode{'}\else$'$\fi}
\newcommand{\arcs}{\ifmmode{''}\else$''$\fi}

\title{The Araucaria Project. First Cepheid Distance to the Sculptor Group Galaxy
NGC 7793 from Variables discovered in a Wide-Field Imaging Survey
\footnote{Based on  observations obtained with the 1.3~m
telescope at the Las Campanas Observatory.
}
}

\author{Grzegorz Pietrzy{\'n}ski}
\affil{Universidad de Concepci{\'o}n, Departamento de Astronomia,
Casilla 160-C,
Concepci{\'o}n, Chile}
\affil{Warsaw University Observatory, Al. Ujazdowskie 4,00-478, Warsaw,
Poland}
\authoremail{pietrzyn@hubble.cfm.udec.cl}
\author{Wolfgang Gieren}
\affil{Universidad de Concepci{\'o}n, Departamento de Astronomia, 
Casilla 160-C, 
Concepci{\'o}n, Chile}
\authoremail{wgieren@astro-udec.cl}
\author{Mario Hamuy}
\affil{Universidad de Chile, Departamento de Astronomia, Casilla 36-D,
Santiago, Chile}
\authoremail{mhamuy@das.uchile.cl}
\author{Giuliano Pignata}
\affil{Departamento de Ciencias Fisicas, Universidad Andres Bello, Avda.
Republica 252, Santiago, Chile}
\authoremail{gpignata@das.uchile.cl}
\author{Igor Soszy{\'n}ski}
\affil{Warsaw University Observatory, Aleje Ujazdowskie 4,
PL-00-478,Warsaw, Poland}
\author{Andrzej Udalski}
\affil{Warsaw University Observatory, Aleje Ujazdowskie 4, PL-00-478,
Warsaw,Poland}
\author{Alistair Walker}
\affil{Cerro Tololo Inter-American Observatory, Casilla 603, La Serena, Chile}
\authoremail{awalker@ctio.noao.edu}
\authoremail{udalski@astrouw.edu.pl}
\author{Pascal Fouqu{\'e}}
\affil{Observatoire Midi-Pyrenees, Laboratoire d'Astrophysique (UMR 5572), 
Universit{\'e} Paul Sabatier, 14, avenue Edouard Belin, Toulouse, France}
\authoremail{pfouque@ast.obs-mip.fr}
\author{Fabio Bresolin}
\affil{Institute for Astronomy, University of Hawaii at Manoa, 2680 Woodlawn 
Drive, Honolulu HI 96822, USA}
\authoremail{bresolin@ifa.hawaii.edu}
\author{Rolf-Peter Kudritzki}
\affil{Institute for Astronomy, University of Hawaii at Manoa, 2680 Woodlawn 
Drive, Honolulu HI 96822, USA}
\authoremail{kud@ifa.hawaii.edu}
\author{Alejandro Garcia-Varela}
\affil{Universidad de los Andes, Carrera 1, N18A, Bogot{\'a}, Colombia}
\authoremail{josegarc@uniandes.edu.co}
\author{Olaf Szewczyk}
\affil{Universidad de Concepci{\'o}n, Departamento de Astronomia,
Casilla 160-C,
Concepci{\'o}n, Chile}
\author{Micha{\l} Szyma{\'n}ski}
\affil{Warsaw University Observatory, Aleje Ujazdowskie 4, PL-00-478,
Warsaw, Poland}
\authoremail{msz@astrouw.edu.pl}
\author{Marcin Kubiak}
\affil{Warsaw University Observatory, Aleje Ujazdowskie 4, PL-00-478,
Warsaw, Poland}
\authoremail{mk@astrouw.edu.pl}
\author{{\L}ukasz Wyrzykowski}
\affil{Warsaw University Observatory, Aleje Ujazdowskie 4, PL-00-478,
Warsaw,Poland}
\affil{Institute of Astronomy, University
of Cambridge, Madingley Road, CB3 0HA, UK}
\authoremail{wyrzykow@astrouw.edu.pl}

\begin{abstract}
We have detected, for the first time, Cepheid variables in the Sculptor Group
spiral galaxy NGC 7793. From wide-field images obtained in the optical V and I
bands on 56 nights in 2003-2005, we have discovered 17 long-period (24-62 days) Cepheids whose
periods and mean magnitudes define tight period-luminosity relations. We use the
(V-I) Wesenheit index to determine a reddening-free true distance modulus
to NGC 7793 of 27.68 $\pm$ 0.05 mag (internal error) $\pm$ 0.08 mag (systematic error). 
The comparison of the reddened distance moduli in V and I with the one derived from
the Wesenheit magnitude indicates that the Cepheids in NGC 7793 are affected by an
average total reddening of E(B-V)=0.08 mag, 0.06 of which is produced inside the host galaxy.

As in the earlier Cepheid studies of the Araucaria Project, the reported distance is tied to 
an assumed LMC distance modulus of 18.50. The quoted systematic uncertainty takes into account 
effects like blending and possible inhomogeneous filling of the Cepheid instability
strip on the derived distance. The reported distance value does not depend on the (unknown)
metallicity of the Cepheids according to recent theoretical and 
empirical results. Our Cepheid distance is shorter, but within the errors consistent 
with the distance to NGC 7793 determined earlier with the TRGB and Tully-Fisher methods. 

The NGC 7793 distance of 3.4 Mpc is almost identical to the one our project had found 
from Cepheid variables
for NGC 247, another spiral member of the Sculptor Group located 
close to NGC 7793 on the sky. 
Two other conspicuous spiral galaxies in the
Sculptor Group, NGC 55 and NGC 300, are  much nearer (1.9 Mpc), confirming the picture of 
a very elongated structure of the Sculptor Group in the line of sight
put forward by Jerjen et al. and others.
\end{abstract}


\section{Introduction}
In our ongoing Araucaria Project (Gieren et al. 2005a), we are engaged in an accurate 
determination of the environmental dependences of several of the most important stellar 
methods of distance determination, with the aim to improve the currently weakest link of the
calibration of the extragalactic distance scale, the determination
of the absolute distances to galaxies in the near field. These
form the fundament for the calibration of secondary methods of distance
determination reaching out into the unperturbed Hubble flow, where the Hubble constant
can be reliably measured from distance and velocity data.
One of the most important stellar methods to measure the distances to 
nearby galaxies
is the period-luminosity (PL) relation obeyed by Cepheid variables. This is particularly
true
when the PL relation is used at near- or mid-infrared wavelengths (e.g. Persson et al. 2004,
Freedman et al. 2008) where the 
effect of reddening on the measured distances becomes small or even negligible.
In the lack of infrared photometry,
an excellent alternative to derive very accurate Cepheid distances is to use
the (V-I) Wesenheit function $W_I$. Recent evidence from our own project
(Pietrzynski and Gieren 2006) and other work (Bono et al. 2010) shows that not only
the effect of reddening on the derived distances is minimized, but also the
effect of the (many times unknown) metallicity of the Cepheid populations in
the target galaxies on the derived distances is very small. In previous papers,
we have reported on the discoveries of abundant Cepheid
populations in a number of irregular galaxies in the Local Group (Pietrzynski et al.
2004, 2006a, 2007), and in several
spiral galaxies members of the Sculptor Group (Pietrzynski et al. 2002, 2006b; Garcia
et al. 2008). These galaxies span a
broad range of metallicities which is useful not only to study the dependence
of the Cepheid PL relation on this parameter, but also of other tools for distance
determination studied in our project, like the Flux-Weighted Gravity-Luminosity Relation
 obeyed by blue supergiant stars
(Kudritzki et al. 2008, Urbaneja et al. 2008).

In this paper, we report on the discovery of a population of Cepheids in another
target galaxy in the Sculptor Group included in our project, NGC 7793. NGC 7793
is a relatively tightly wound spiral of type Sd (Sandage \& Bedke 1988), similar in appearance
to NGC 300 in the Sculptor Group. Unlike NGC 300 however, little work has been
done in the past to determine the distance to NGC 7793. Only two distance indicators have been 
applied on NGC 7793: the I-band TRGB method, and the Tully-Fisher (TF) method. While the
TF method has yielded a distance modulus of 28.06 $\pm$ 0.35 mag (Tully et al. 2009),
application of the TRGB technique has resulted in a distances of 27.96 $\pm$ 0.24 mag
(Karachentsev et al. 2003), and more recently of 27.79 $\pm$ 0.08 mag (Jacobs et al. 2009).
The Jacobs et al. determination was made from HST data obtained in an outer field of NGC 7793
which explains its higher accuracy, as compared to the result of Karachentsev et al. which
was measured from HST data obtained for a field located much closer to the center of the galaxy, 
where crowding and internal extinction may have complicated the TRGB measurement.
Given that NGC 7793 contains a considerable number of blue massive stars indicating
recent star formation, we expected to find a sizeable population of classical Cepheids
in it which would provide an independent and accurate check on the previous distance
determinations. Indeed, our survey has led to the detection of 17 long-period Cepheids.
In this paper, we are using these Cepheids to determine the first Cepheid-based distance 
to NGC 7793, which adds a valuable target to the existing Araucaria Project database
with a distance from this indicator.

Our paper is organized as follows.
In section 2, we describe our observations, data reductions and calibrations.
In section 3, we present the catalog of photometric properties of the Cepheid variables
discovered in our survey in NGC 7793. In section 4, we construct the period-luminosity
relations in V, I and the reddening-free (V-I) Wesenheit index and determine the distance
and internal reddening
of NGC 7793 from these data. In section 5, we discuss our result and assess its accuracy.
The main conclusions are presented in section 6.

\section{Observations,  Reductions and Calibrations}
All the  data presented in this paper were collected with the Warsaw 1.3-m 
telescope at Las Campanas Observatory. The telescope was equipped with 
a mosaic 8k $\times$ 8k detector, with a field of view of about 35 $\times$ 35 
arcmin and a scale of about 0.25 arcsec/pix. For more  instrumental
details on this camera, the reader is referred to the OGLE  website:
{\it http://ogle.astrouw.edu.pl}.
V and I band  images of NGC 7793  were secured on 56 different nights.
The exposure time was set to 900 seconds in both filters.
The observations were obtained between July 2003 and December 2005.

Preliminary reductions (i.e. debiasing and flatfielding)  were 
done with the IRAF\footnote{IRAF is distributed by the
National Optical Astronomy Observatories, which are operated by the
Association of Universities for Research in Astronomy, Inc., under cooperative
agreement with the NSF.} package. Then, the PSF photometry was obtained 
for all stars in the same manner as in Pietrzy{\'n}ski, Gieren and
Udalski (2002). Independently, the data were reduced with the OGLE III pipeline 
based on the image subtraction technique (Udalski 2003; Wo{\'z}niak 2000).  

In order to calibrate our photometry onto the standard system 
our target was monitored during two (non-contiguous) photometric nights together with 
some 25 standards from the Landolt fields spanning a wide range of
colors ( -0.14 $<$ V-I $<$ 1.43), and observed at widely different
airmasses. Since in principle the transformation equations for each of the eight
chips may have different color coefficients and zero points, the selected
sample of standard stars was observed on each of the individuial chips, and
transformation coefficients  were derived independently for each chip, 
on each night. Comparison of the photometry obtained on the different nights
revealed that the internal accuracy of the zero points in both V and
I bands is better than 0.03 mag. In order to demonstrate this, we show
in Table 3 the zero point and color
term values for the eight chips for the two calibrating photometric
nights. Within the errors there are no color term variations from night
to night. Slight zero point differences for individual chips from 0.02-0.04 mag
are present. We note here this (OGLE) telescope and setup has been used now for
14 years for photometric surveys and is certainly one of the most stable
photometric systems in the world. Never any significant color drift has been
observed.

To correct the variation of the zero points in V and I
over the mosaic, the "correction maps" established by Pietrzynski 
et al. (2004) were used. These maps were already applied to correct 
photometry obtained in the field of NGC 6822 (Pietrzynski et al. 
2004) and NGC 3109 (Pietrzynski et al. 2006a). Comparison with 
other studies revealed that these maps allow to correct 
the zero point variations down to a level of 0.02-0.03 mag.

Unfortunately, we were unable to find any  modern photometry of 
of NGC 7793 in the literature which could be used to check the
accuracy of the zero point of our photometry.
However, taking into account that the present photometric calibration 
was performed in an identical way as for the other nearby galaxies 
surveyed for Cepheids in the Araucaria project, and that in all cases a comparison 
of our photometry with the corresponding results published in the literature
had shown very good agreement,
we are very confident that  our photometric data of NGC 7793  are also accurate to 0.03 mag 
in both V and I, and that a possible variation of the photometric zero points in both filters
 over the mosaic fields is indeed smaller than 0.03 mag.

\section{Cepheid Catalog}
All stars detected in NGC 7793 were searched for photometric variations with periods between
0.2 and 200 days, using the analysis-of-variance algorithm (Schwarzenberg-Czerny 1989).
In order to distinguish Cepheids from other types of variables, we used the same
criteria given in our initial paper reporting on the discovery of Cepheids in NGC 300
(Pietrzynski et al. 2002). The light curves of all Cepheid candidates were approximated
by a 4th order Fourier series. We then rejected objects with amplitudes of their
V light curves smaller than 0.4 mag, which is the approximate lower limit of V amplitudes
for classical Cepheids. This procedure also helped us in screening our Cepheid sample
from the inclusion of strongly blended variables. For the variables passing our selection
criteria, the mean V and I magnitudes were derived by integrating their light curves
which had been previously converted onto an intensity scale, and converting the results
back to the magnitude scale. 

Our final catalog contains 17 classical Cepheids with periods between 24 and 62 days.
Fourteen of these variables have light curves in both V and I filters. Three additional
Cepheids have only V-band light curves. Two of them are the faintest Cepheids in our
sample and were too faint to measure their I-band light curves from our images. The third
Cepheid with V-band data only is the longest-period variable we discovered (91 days),
but unfortunately it was located in a gap between the chips of our mosaic camera 
in the I images. There are
certainly many more Cepheids in NGC 7793 with shorter pulsation periods which we
were not able to discover because of the small size of the telescope we had at our
disposal for this survey; yet, the sample of long-period and relatively bright Cepheids
detected in our survey is completely adequate for our purpose to derive an accurate
distance to NGC 7793, as we will show in the next section.

The journal of the individual 
V and I observations of the Cepheids is presented in Table 1. Table 2 summarizes the photometric
properties of the newly discovered Cepheids in NGC 7793: their periods, and their intensity
mean magnitudes in V, I and in the (V-I) Wesenheit band (defined as $W_I$ = I - 1.55($<V> -
<I>$); see Udalski et al. 1999). The
periods of the Cepheids are typically accurate to $10^{-3}*P$. Table 2 also reports 
precise equatorial coordinates for all the variables. In Figure 1, we show the locations of the
Cepheids discovered in our survey in their host galaxy NGC 7793.

In Figure 2, we show the light curves of the complete sample of NGC 7793 Cepheids 
which illustrate the good quality and phase coverage of our data. 
Figure 3 shows the locations of the Cepheids on the V, V-I 
color-magnitude diagram for NGC 7793 constructed from our data. It is seen that all Cepheids
lie in the expected region of the CMD which delineates the Cepheid instability strip,
presenting supporting evidence that all variables in Table 2
were classified correctly as classical
Cepheids. Figure 3 does of course not include the 3 Cepheids which have no I-band light
curves, but the shapes and amplitudes of their V light curves, periods and mean V
brightnesses leave no doubt that they are classical Cepheids too.

\section{PL relations and distance determination}
In Figures 4-6, we show the PL relations in the V, I and $W_I$ bands resulting from the data
in Table 2.  The V-band PL diagram contains the 3 variables which have V data only; these
3 variables are, however, not included in the distance determination.
Fits to a straight line to our data yield the following 
slopes for the PL relations: -3.24 $\pm$ 0.25, -3.17 $\pm$ 0.18 
and -3.07 $\pm$ 0.30 in V, I and ${\rm W_{\rm I}}$, respectively. The uncertainties on
the slopes are rather large due to the relatively small number of Cepheids in the PL
diagrams. The slope values however
are consistent, at a level of 2 $\sigma$,  with the corresponding OGLE slopes of -2.775, -2.977 
and -3.300 for the LMC (Udalski 2000). We therefore use, as in our previous papers, the
extremely well defined OGLE PL relation slopes determined for the LMC to fit our observed
PL diagrams in NGC 7793. Remarks on the validity of this approach will be given in the
next section.  Adopting the LMC slopes in fitting our data leads to the following 
equations:\\

V = -2.775 log P + (26.48 $\pm$ 0.04) \\

I = -2.977 log P + (25.91 $\pm$ 0.03) \\

${\rm W}_{\rm I}$ = -3.300 log P + (25.04  $\pm$ 0.04) \\

The uncertainties on the zero points have been calculated from the least-squares
fitting.
Adopting 18.50 mag as the true distance modulus to the LMC, as we did in our previous
papers in this series, a value of
E(B-V) = 0.018 mag as the Galactic foreground 
reddening towards NGC 7793 (Schlegel et al. 1998) and the reddening law given by
the same authors ( ${\rm A}_{\rm V}$ = 3.24 E(B-V), ${\rm A}_{\rm I}$ = 1.96
E(B-V)) we derive the following distance moduli for NGC 7793 in the
three different bands:\\

$(m-M)_{0}$ (${\rm W}_{\rm I}$) = 27.68 $\pm$ 0.05 mag \\

$(m-M)_{0}$ (I) = 27.78 $\pm$ 0.03 mag\\

$(m-M)_{0}$ (V) = 27.86 $\pm$ 0.04 mag\\

Again, the uncertainties on these values are derived from least-squares fitting.
The distance moduli values indicate, not surprisingly, that there is additional reddening 
intrinsic to NGC 7793
affecting the Cepheids. A value of E(B-V)=0.06 mag for the average intrinsic
reddening, bringing the total mean reddening affecting the NGC 7793 in our sample
to 0.08 mag, produces excellent agreement, at the 0.02 mag level, between the
distance moduli for NGC 7793 derived from the PL relations in V, I and ${\rm W}_{\rm I}$. 
We adopt, as our best Cepheid distance from the present optical study, the value of 27.68 mag
derived from the Wesenheit index. In the next section, we discuss the various systematic
uncertainties which may affect this value, and estimate the total uncertainty
on this first Cepheid distance determination to NGC 7793.

\section{Discussion}
The current distance measurement of NGC 7793 is the first one based on Cepheid variables,
and subject to the different sources of systematic uncertainty inherent to this particular
method. The first potential problem is with reddening because Cepheids, as relatively
young stars, tend to be located in dusty regions in their host galaxies. While infrared 
photometry is the best way to minimize or even completely avoid this problem, our past work, 
and that of others, most notably the OGLE Project (Udalski et al. 1999; Soszynski et al. 2008),
has shown that in Cepheid distance work reddening can be very effectively dealt with by
using the (V-I) Wesenheit magnitude, as we have done in this paper. In all our previous
papers in this series, the true distance moduli based on the Cepheid Wesenheit magnitudes agreed
within the (small) uncertainties of the determinations with the distances coming from the
near-infrared follow-up studies (e.g. Gieren et al. 2005b, Gieren et al. 2008, 2009). While
our intention is to check on the current distance determination to NGC 7793 
using near-infrared data
in the near future, we believe that the Cepheid distance presented in this paper is
basically free of any significant error due to inadequate absorption corrections. In any
case, NGC 7793 is a favorable target in this sense given that the Galactic foreground
reddening towards the galaxy is practically negligible, and from our results in the
previous section we know that the average reddening produced inside the galaxy on the
Cepheids is quite small too.

At a distance of 3.4 Mpc blending of Cepheids with nearby stars not resolved in the
photometry is certainly a significant problem. For one of our target galaxies, NGC 300,
we have made a stringent test on the effect of blending on a Cepheid-based distance,
comparing the results from ground-based photometry in V and I to HST-based photometry
of the same Cepheids (Bresolin et al. 2005). The result was that blending affects the
ground-based distance result by less than 2\%, in this particular case. NGC 7793 has a
very similar inclination with respect to the line-of-sight, with an orientation close
to face-on, as NGC 300; however, since its distance is by a factor of 1.8 larger, the effect of
blending is expected to be more serious than in the case of NGC 300. Scaling linearly
with the distance,
we expect that 3\% is a reasonable upper limit to the systematics from this source. We
tested this hypothesis by degrading the resolution of the ground-based images of NGC 300
to the one expected at the distance of NGC 7793, and redoing the photometry of the Cepheids
observed with HST. The result is that indeed the effect of blending is somewhat larger now, 
but less than 3\%, in agreement with our simple expectation from rescaling the error
according to the larger distance of NGC 7793. It should also
be noted in this context
that the amplitudes of the light curves in Figure 1 are generally about as large
as expected for the periods of the Cepheids (although there is no strict period-amplitude
relation for classical Cepheids), which probably implies that none of these Cepheids
is strongly blended, which depending on the color of a bright nearby star contaminating
the Cepheid photometry would tend to reduce the light curve amplitude in V (blue companion)
or I (red companion) (Gieren 1982). Regarding the effect of blending,
it should be recalled that it acts to make Cepheids brighter, and thus to 
underestimate the distance to their host galaxy. 

Since the PL relation method is a statistical one and requires that the PL plane is
populated by a sufficient number of Cepheid variables to minimize problems due to an
incomplete, selective filling of the instability strip, the relatively small number of
14 Cepheids in our present study is a potential reason of concern. However, both the
filling of the instability strip on the CMD in Figure 2 by the variables in our catalog,
as well as the non-systematic scatter of the residuals around the fitting lines in Figures
3-5 suggest that the Cepheid variables in this study do populate the instability strip
quite homogeneously, with no clear preference towards either the blue nor the red edge of the
strip. Also, the current distance determination should not be affected at a significant
level by a Malmquist bias which could originate when the faintest Cepheids in the sample
were very close to the cutoff magnitude of the photometry. Figure 2 suggests that the
faintest of the 14 Cepheids used in the distance determination is still about a full
magnitude brighter than the faint star limit of our photometry. The effect of a Malmquist
bias is to make the slope of the observed PL relation too shallow because
at the short-period (faint) end of the sample only Cepheids are detected which
are very bright, for the value of their period. Another argument giving us confidence
that our distance determination is not significantly affected by a Malmquist bias, is
the fact that the slopes of the free linear fits to a line in the V- and I-band PL planes 
are both steeper than the assumed true (LMC) slopes on these diagrams, which is the opposite
to what would be expected if a significant Malmquist bias was indeed affecting the data in
Figures 3 and 4.

A problem which has been very intensively discussed in the recent literature is the effect
of metallicity on the Cepheid PL relation, both in optical and near-infrared photometric
bands. A very nice and complete summary of the work on the metallicity effect on the PL
relation, and corresponding conclusions both from observations and theoretical modelling
of Cepheids, has been recently presented by Bono et al. (2010). In particular, these authors
find that the (V-I) Wesenheit PL relation used in this study to determine the distance
to NGC 7793 has a slope which {\it independently of metallicity} agrees always very well
with the LMC slope we have been using in our analysis. This empirical result is backed up
by the predictions of their pulsation models for fundamental mode Cepheids and thus appears
to be a very solid conclusion. In addition, Bono et al. conclude from their analysis that
the {\it zero point} of the absolute (V-I) Wesenheit PL relation does not depend on metallicity
either, over a broad range of metallicities, a conclusion reached from the available data
and again supported by their theoretical models. We note here that from the (preliminary)
dataset from the Araucaria Project alone available to us in 2006, we had reached the same
conclusion (Pietrzynski \& Gieren 2006). As a conclusion, any effect of the (currently
unknown) mean metallicity of the Cepheids in our NGC 7793 sample on the distance
we have derived in this paper, should be negligible.

An issue of concern is the fact that the observed dispersion in the $W_I$ Wesenheit PL relation
in Figure 6 is slightly larger than the one shown by the I-band PL relation in Figure 5. We believe
that the main reason for this is the photometric contamination of some of the Cepheid mean magnitudes
in Table 2 by nearby, unresolved companion stars (discussed above). While this effect indeed tends to 
increase the scatter in Figure 6, it should not change the distance derived from this diagram within
the uncertainties we determined in our analysis in section 4 of this paper. An alternative
explanation would be the validity of a non-standard reddening law in NGC 7793, the coefficient 1.55 
not being
appropriate in the definition of the Wesenheit magnitude to correct for reddening. We regard this
as rather unlikely however, in the light of our earlier findings in NGC 300 (Gieren et al. 2005b),
NGC 55 (Gieren et al. 2008) and NGC 247 (Gieren et al. 2009) where from combined optical and
near-infrared photometry of the Cepheids we found solid evidence that the standard reddening law
holds in these galaxies.

 From the previous discussion, we identify as the dominant systematic uncertainty in the
present study on the derived distance the effect of unresolved bright nearby stars
on the Cepheid photometry, or blending. The systematic uncertainty on the photometric zero points 
in our study are less important (less than 2\% on the distance modulus). We estimate the total
systematic uncertainty of our distance to be 4\%, with a likely tendency to act in the
direction to increase the reported distance to NGC 7793. This would bring the Cepheid
distance closer to the TRGB-based distance of Jacobs et al. (2009), which are clearly
consistent within their respective uncertainties.

We recall that the current distance measurement of NGC 7793 {\it assumes} an LMC distance
of 18.50 mag, in agreement with our earlier work in the Araucaria Project. While any possible 
future change in the adopted value of the LMC distance will obviously affect the absolute
Cepheid distances of our different target galaxies
reported in the Araucaria project, it will not affect the
{\it relative} distances between the target galaxies of our project.

\section{Conclusions}
We have carried out a first systematic wide-field search for Cepheid variables in the Sculptor 
Group spiral NGC 7793 and have discovered 17 long-period classical Cepheids. We provide the
periods of these objects and their mean intensity magnitudes in the optical V and I bands,
as well as their accurate positions. From a subsample of 14 Cepheids having both V- and I-band 
light curves, we have determined the period-luminosity relations in these bands, as well as 
in the reddening-independent (V-I) Wesenheit band. The distance moduli derived from these PL
relations demonstrate that apart from the very small foreground reddening, there is additional
reddening of magnitude E(B-V)=0.06 mag produced in NGC 7793 itself and affecting its
Cepheids. Our best adopted value for the distance modulus of NGC 7793 is 27.68 $\pm$ 0.05 mag
(intrinsic) $\pm$ 0.08 mag (systematic). The dominant source of systematic uncertainty is the
effect unresolved nearby bright stars might have on the photometric magnitudes of some of the
Cepheids, which would act towards a (small) underestimation of the true distance of the galaxy. 
According to recent
studies, our way to determine the distance to NGC 7793 using the period - (V-I) Wesenheit magnitude
diagram should be independent of the metallicities of the Cepheids in our sample. 

The Cepheid distance agrees very well with the recent distance determination of NGC 7793
of Jacobs et al. (2009) who used HST photometry and the I-band TRGB method. This increases
our confidence that the distance to NGC 7793 is now reliably measured with a total
uncertainty not exceeding 5\%. Nevertheless, we intend to obtain near-infrared 
single-phase photometry for our Cepheid sample with the ESO VLT to determine accurate
mean magnitudes in the J and K bands (see Soszynski et al. 2005) and use our multiwavelength
approach (Gieren et al. 2005b) to improve the accuracy of our present distance determination
from optical photometry of the Cepheids.

The present study confirms the elongated structure of the Sculptor Group in the line-of-sight
first suspected by de Vaucouleurs (1959),  and later substantiated by Jerjen et al. (1998),
with NGC 55 and NGC 300 at the near end with near-identical distances of 1.9 Mpc,
and NGC 247 and NGC 7793 at the far end at about 3.5 Mpc. From the Cepheid work in the
Araucaria Project, the distances to these four galaxies are now all determined to 5\%
or better.

\acknowledgments
We are grateful to Las Campanas 
Observatory, and to the CNTAC for providing the large amounts of 
telescope time which were necessary to complete this project.
We gratefully acknowledge financial support for this
work from the Chilean Center for Astrophysics FONDAP 15010003, and from
the BASAL Centro de Astrofisica y Tecnologias Afines (CATA) PFB-06/2007.
Support from the Polish grant N203 387337 and the FOCUS
subsidy of the Fundation for Polish Science (FNP) is also acknowledged.
PF suddenly feels old realizing that he started his astronomical work measuring
the distance of NGC 7793 through the H I Tully-Fisher relation in 1978, finding
a similar distance to the one we report in this paper. We would also like to thank the 
anonymous referee for his comments.

Last not least, WG is grateful to Conicyt, to the Astronomy Department at the
Universidad de Chile, and to the Sociedad Chilena de Astronomia SOCHIAS for their
encouragement and financial aid which helped the Astronomy Department at the
Universidad de Concepci{\'o}n to quickly recover from the material damages
suffered from the earthquake which struck us on February 27, 2010.

\begin{deluxetable}{ccccc}
\tablecaption{Individual V and I Observations}
\tablehead{
\colhead{object}  & \colhead{filter} &
\colhead{HJD-2450000}  & \colhead{mag}  & \colhead{$\sigma_{mag}$}\\
}
\startdata
cep001 & V & 2851.81845 &  20.744 &   0.040 \\
cep001 & V & 2853.88401 &  20.762 &   0.034 \\
cep001 & V & 3260.81098 &  21.247 &   0.066 \\
cep001 & V & 3263.80269 &  21.113 &   0.048 \\
cep001 & V & 3267.79044 &  21.168 &   0.050 \\
cep001 & V & 3270.82209 &  21.207 &   0.065 \\
cep001 & V & 3286.68579 &  20.913 &   0.051 \\
cep001 & V & 3289.69621 &  20.913 &   0.049 \\
cep001 & V & 3295.66265 &  20.804 &   0.038 \\
cep001 & V & 3297.65958 &  20.683 &   0.033 \\
cep001 & V & 3299.68023 &  20.733 &   0.059 \\
cep001 & V & 3305.61427 &  20.666 &   0.086 \\
cep001 & V & 3311.63347 &  20.868 &   0.043 \\
cep001 & V & 3314.66526 &  20.859 &   0.052 \\
cep001 & V & 3324.66691 &  21.055 &   0.058 \\
cep001 & V & 3327.65426 &  21.099 &   0.078 \\
cep001 & V & 3328.53947 &  20.924 &   0.076 \\
cep001 & V & 3339.53311 &  21.110 &   0.083 \\
cep001 & V & 3354.55446 &  21.253 &   0.066 \\
cep001 & V & 3615.70979 &  20.985 &   0.045 \\
\enddata
\tablecomments{The complete version of this table is in the electronic
edition of the Journal.  The printed edition contains only
the the first 20 measurements in V band for the Cepheid variable cep001.}

\end{deluxetable}

\begin{deluxetable}{c c c c c c c c}
\tablecaption{Cepheids in NGC 7793}
%\tablewidth{0pt}
\tablehead{
\colhead{ID} & \colhead{RA} & \colhead{DEC}  & \colhead{P} & \colhead{ ${\rm
T}_{0}$ - 2,450,000} &
\colhead{$<V>$} & \colhead{$<I>$} & \colhead{$<W_{\rm I}>$}\\
 & \colhead{(J2000)} & \colhead{(J2000)}  &
\colhead{ [days]} &  &
\colhead{[mag]} & \colhead{[mag]} & \colhead{[mag]}
}
\startdata
cep001 & 23:57:56.89 & -32:34:16.1 & 91.5667 & 3639.6604 &  20.992 & --- &    --- \\
cep002 & 23:58:03.73 & -32:36:01.1 & 62.1195 & 3654.5911 &  21.484  & 20.469 & 18.896  \\
cep003 & 23:58:01.63 & -32:33:47.8 & 57.6004 & 3615.7098 &  21.413  & 20.676 & 19.534 \\
cep004 & 23:57:35.02 & -32:35:36.1 & 54.7885 & 2574.6127 &  21.458  & 20.679 & 19.472 \\
cep005 & 23:57:57.03 & -32:36:37.2 & 54.0103 & 3615.7098 &  21.448  & 20.606 & 19.301 \\
cep006 & 23:57:58.79 & -32:36:23.3 & 48.0077 & 3267.7904 &  21.748  & 20.844 & 19.443 \\
cep007 & 23:58:08.96 & -32:36:12.3 & 47.7829 & 2497.9038 &  22.024  & 21.066 & 19.581 \\
cep008 & 23:57:54.66 & -32:35:44.1 & 39.7409 & 2560.6109 &  22.150  & 21.269 & 19.903 \\
cep009 & 23:57:48.31 & -32:34:21.3 & 36.4671 & 2499.8403 &  22.230  & 21.260 & 19.757 \\
cep010 & 23:58:06.10 & -32:34:20.3 & 31.2549 & 3299.6802 &  22.526  & 21.514 & 19.945 \\
cep011 & 23:58:01.40 & -32:36:23.8 & 28.0112 & 3260.8110 &  22.437  & 21.631 & 20.382 \\
cep012 & 23:58:11.50 & -32:36:35.1 & 27.6007 & 3684.5384 &  22.433  & 21.527 & 20.123 \\
cep013 & 23:57:47.65 & -32:38:00.4 & 26.3644 & 2845.6831 &  22.517  & 21.597 & 20.171 \\
cep014 & 23:58:00.35 & -32:39:19.6 & 26.2164 & 3626.6284 &  22.757  & 21.788 & 20.286 \\
cep015 & 23:57:47.02 & -32:37:14.4 & 26.0831 & 2497.8846 &  22.550  & 21.785 & 20.599 \\
cep016 & 23:58:09.23 & -32:33:41.1  & 25.4582 & 2641.5723 &  22.758  & ---    &  --- \\
cep017 & 23:57:28.61 & -32:34:18.3 & 24.3677 & 3281.7234 &  22.682  & ---    &  --- \\
\enddata
\end{deluxetable}

\begin{deluxetable}{c c c c c}
\tablecaption{The color terms ($\alpha$)              
and zero points ($\beta$) for chips 1-8 for the two calibrating photometric nights}
\tablehead{
\colhead{chip} & \colhead{$\alpha_{N1}$} & \colhead{$\beta_{N1}$}  & \colhead{$\alpha_{N2}$} 
& \colhead{$\beta_{N2}$} \\
}
\startdata
1  &  -0.025 &  -2.207 &  -0.027 &  -2.239 \\
2  &  -0.018 &  -2.193 &  -0.014 &  -2.201 \\
3  &  -0.020 &  -2.204 &  -0.024 &  -2.233 \\
4  &  -0.020 &  -2.195 &  -0.013 &  -2.210 \\
5  &  -0.017 &  -2.123 &  -0.029 &  -2.159 \\
6  &  -0.029 &  -2.173 &  -0.019 &  -2.223 \\
7  &  -0.010 &  -2.179 &  -0.022 &  -2.204 \\
8  &  -0.024 &  -2.197 &  -0.022 &  -2.137 \\
\enddata
\end{deluxetable}

\begin{figure}[htb]
\vspace*{22cm}
\includegraphics{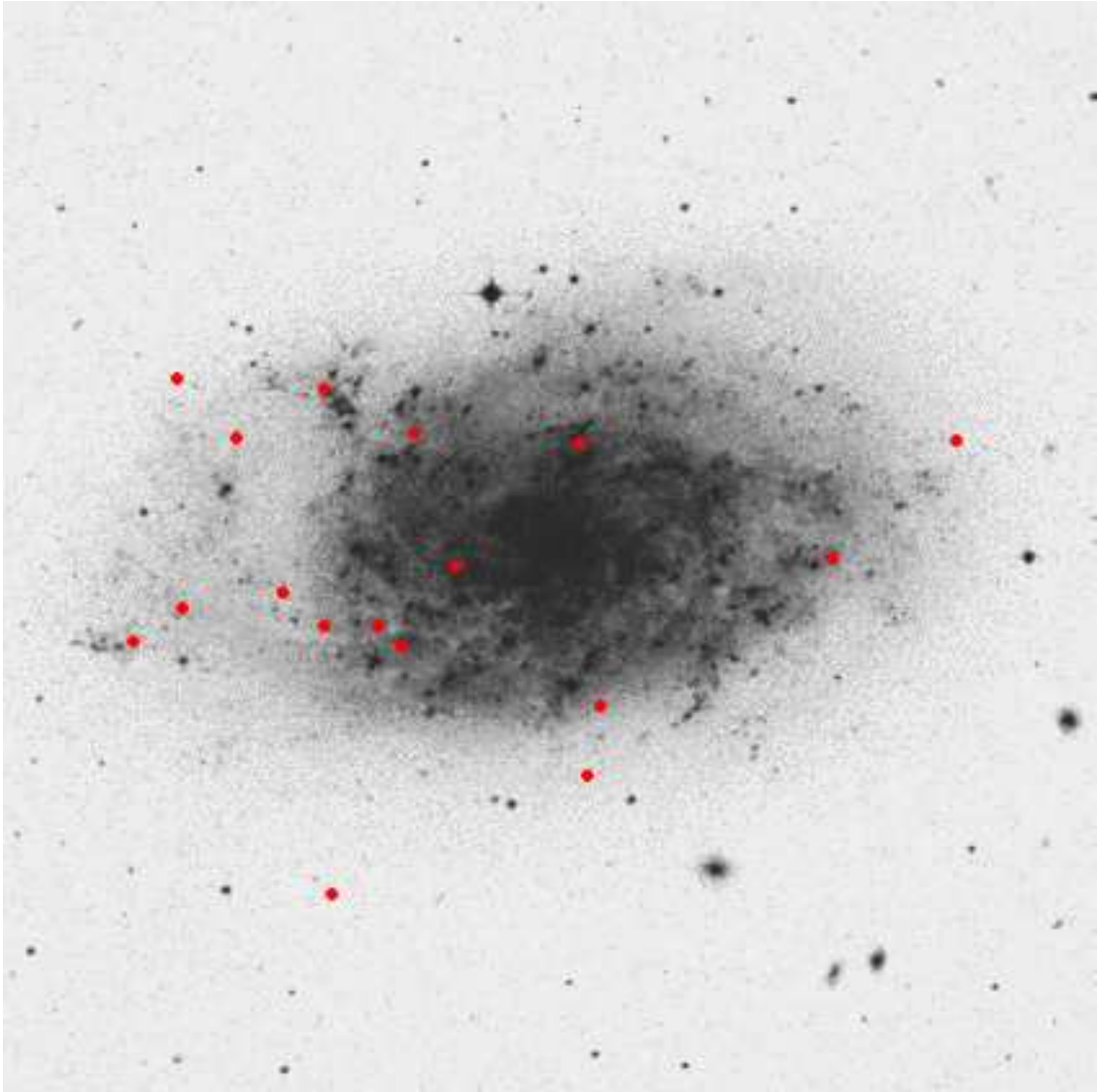}
\caption{
DSS map of NGC 7793, showing the positions of the discovered Cepheids. 
North is up, and east to the left. The field of view is about 12' x 12'
}
\end{figure}

\begin{figure}[htb]
\vspace*{22cm}
\includegraphics{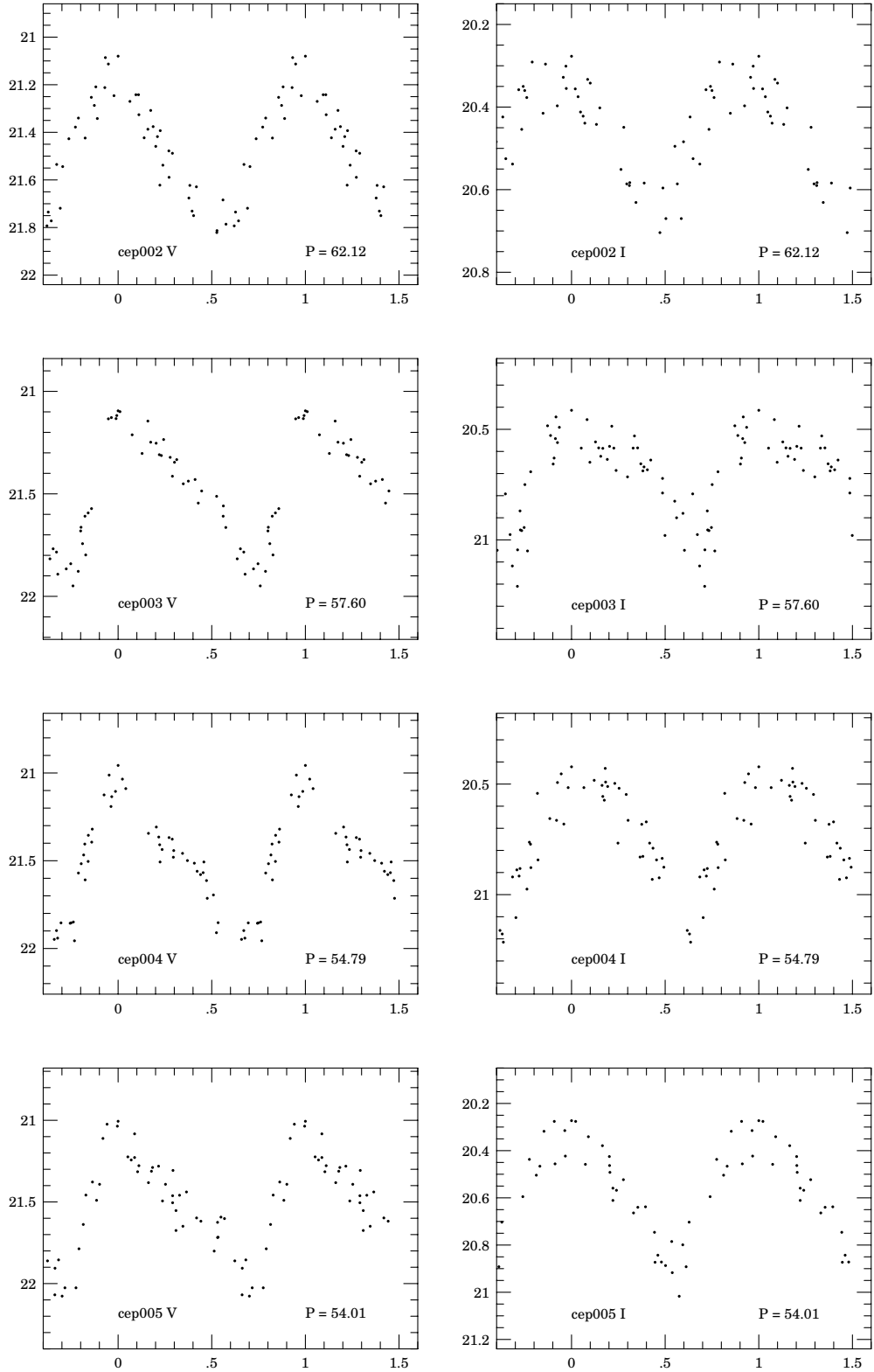}
\caption{The V- and I-band light curves for the 17 NGC 7793 Cepheids 
discovered in our survey.
Individual observations are listed in Table 1, and
periods to phase the observations were taken from Table 2.
}
\end{figure}

\setcounter{figure}{1}

\begin{figure}[htb]
\vspace*{22cm}
\includegraphics{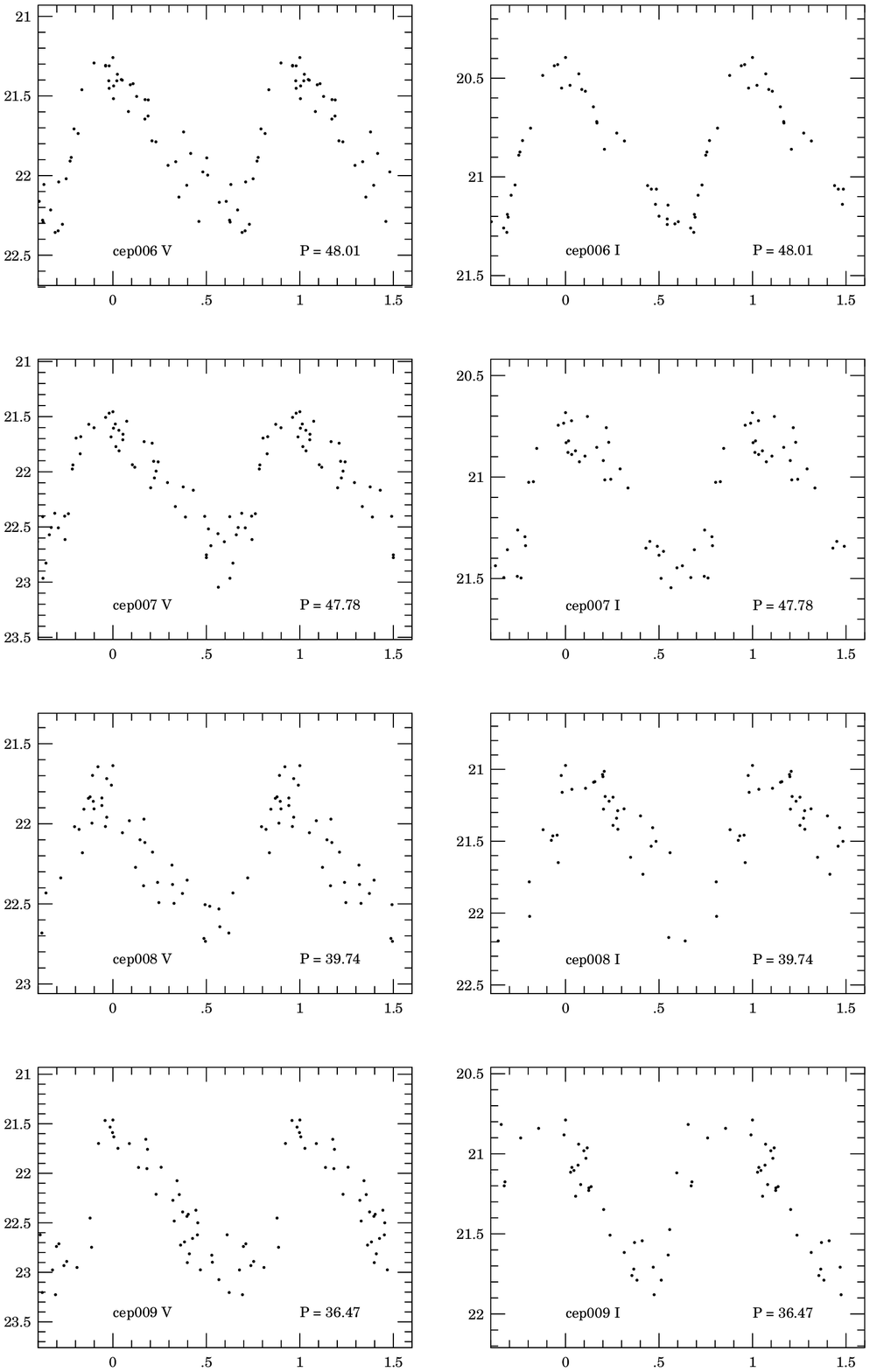}
\caption{Continuation}
\end{figure}

\setcounter{figure}{1}

\begin{figure}[htb]
\vspace*{22cm}
\includegraphics{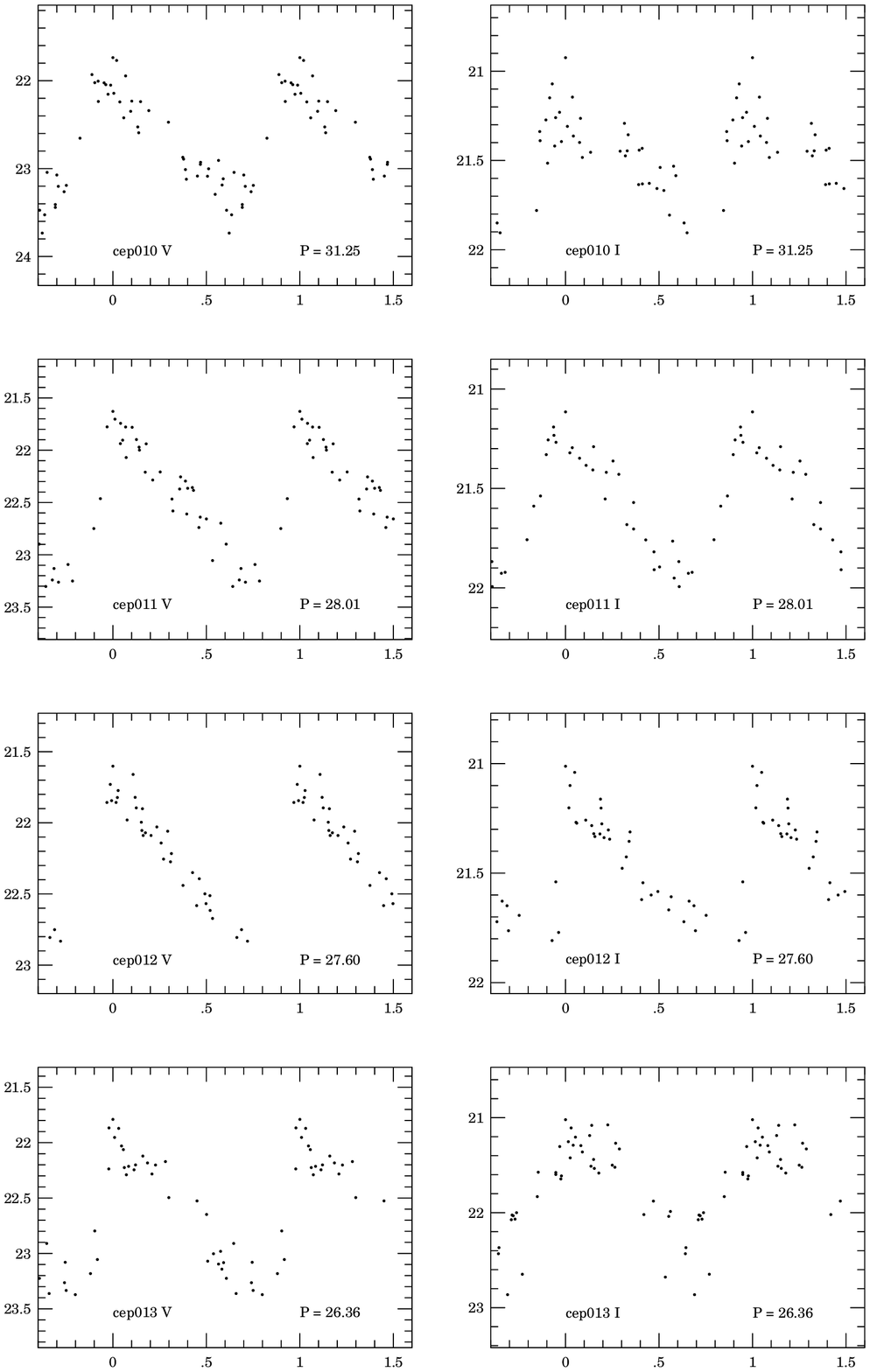}
\caption{Continuation}
\end{figure}

\setcounter{figure}{1}

\begin{figure}[htb]
\vspace*{22cm}
\includegraphics{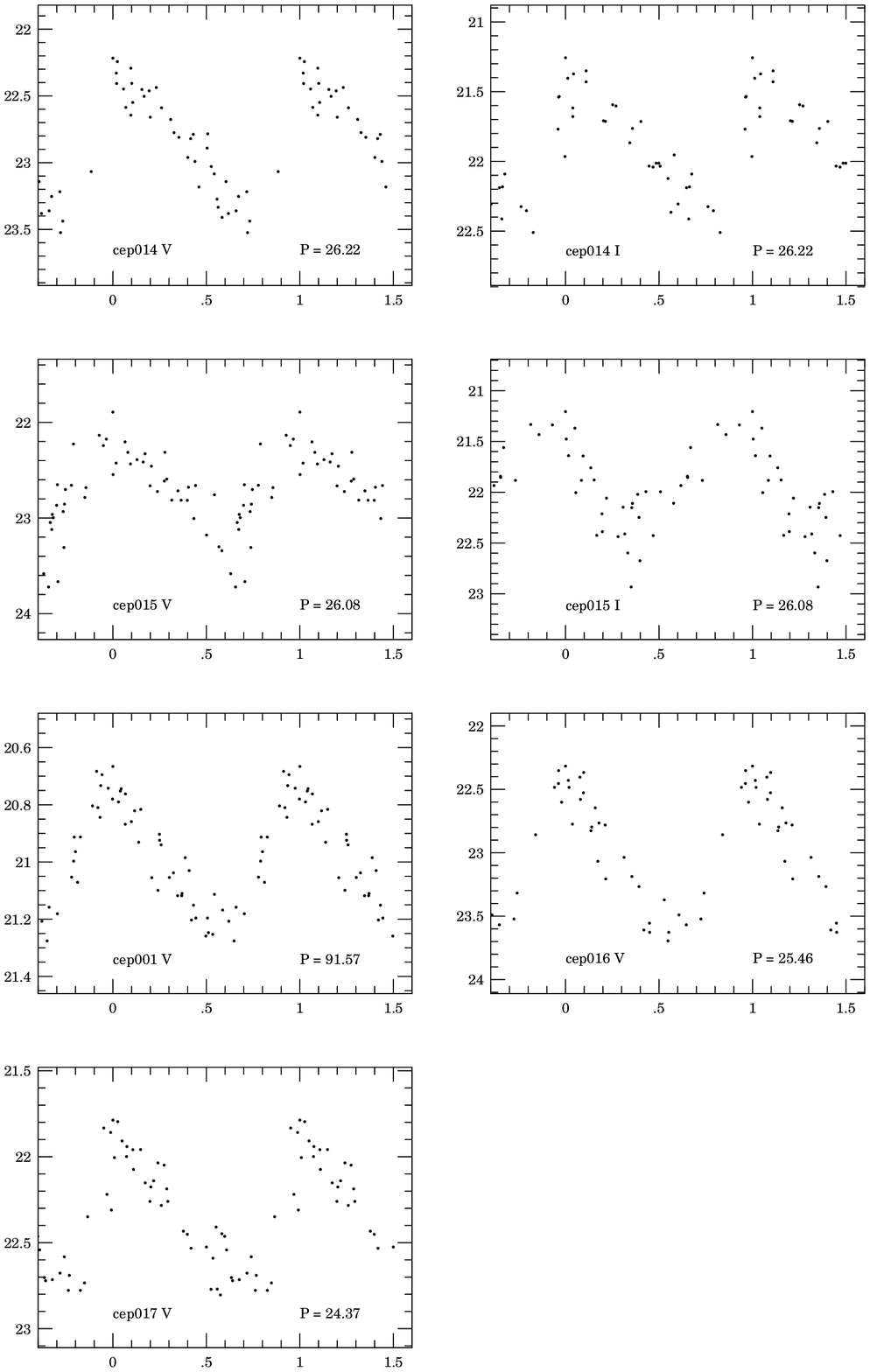}
\caption{Concluded}
\end{figure}

\begin{figure}[htb]
\vspace*{15cm}
\includegraphics{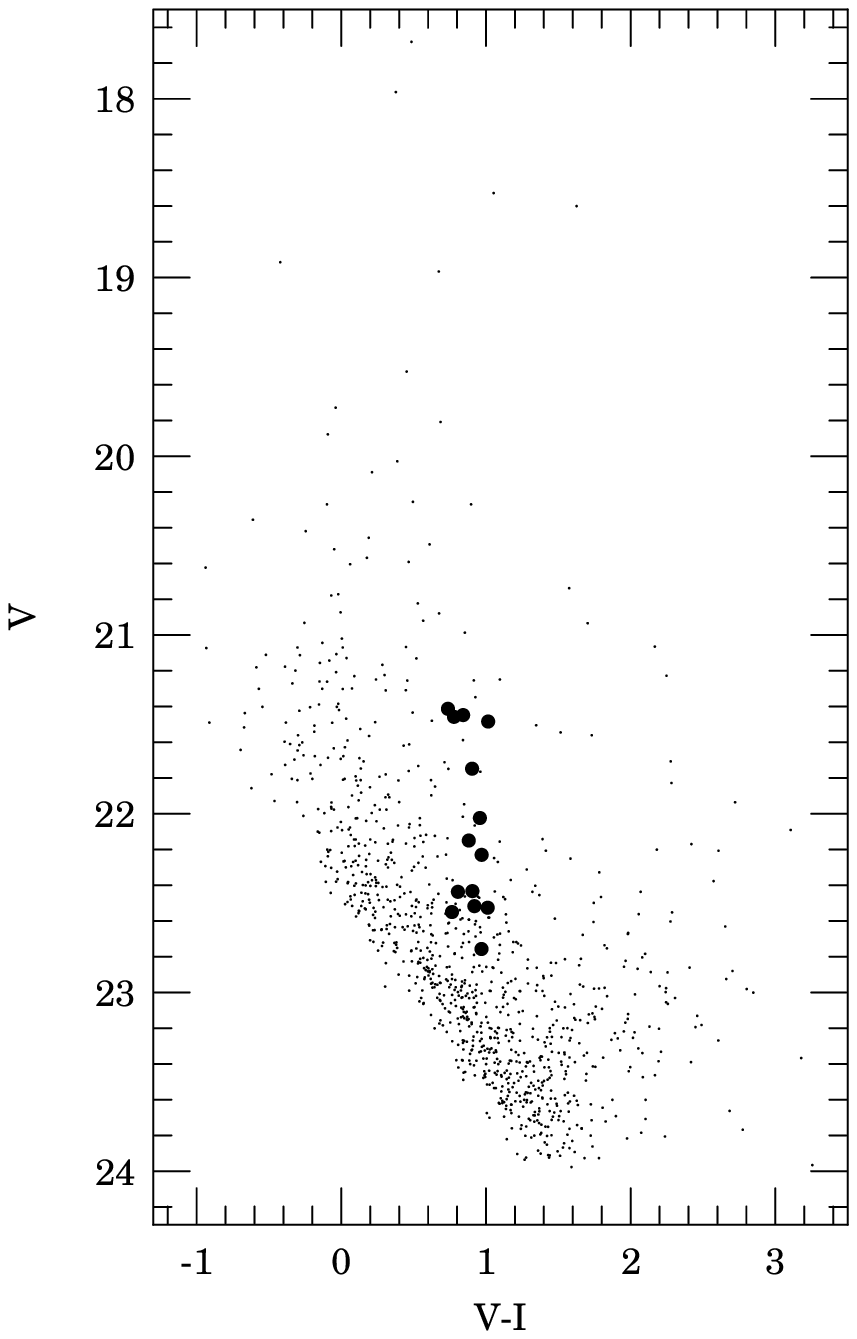}
\caption{The V,V-I magnitude-color diagram for NGC 7793. The discovered
Cepheids (filled circles) populate the expected instability strip for classical Cepheids
in this diagram. It is seen that the faintest Cepheid in our sample is still about
1 mag brighter than the faintest stars in NGC 7793 detected and measured in our
photometry, eliminating a Malmquist bias as a significant source of error in our
distance determination.
}
\end{figure}

\begin{figure}[htb]
\vspace*{15cm}
\includegraphics{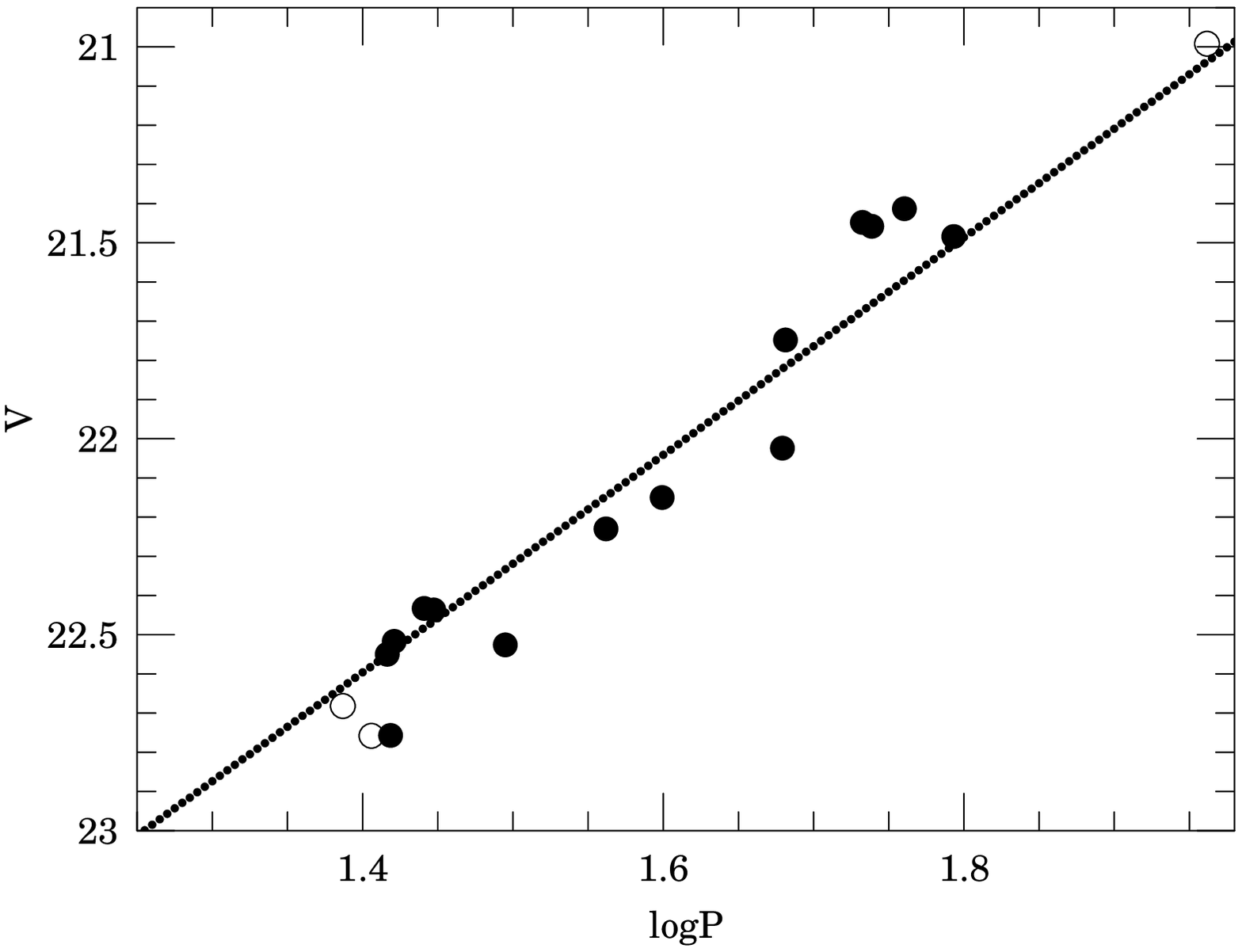}
\caption{The period-luminosity relation for NGC 7793 Cepheids in the V 
band. The 14 Cepheids indicated with filled circles have both V and I mean
magnitudes. The 3 Cepheids marked with open circles have only V-band data.
The slope of the fitting line was adopted from the LMC Cepheids (OGLE II).
}
\end{figure}

\begin{figure}[htb]
\vspace*{15cm}
\includegraphics{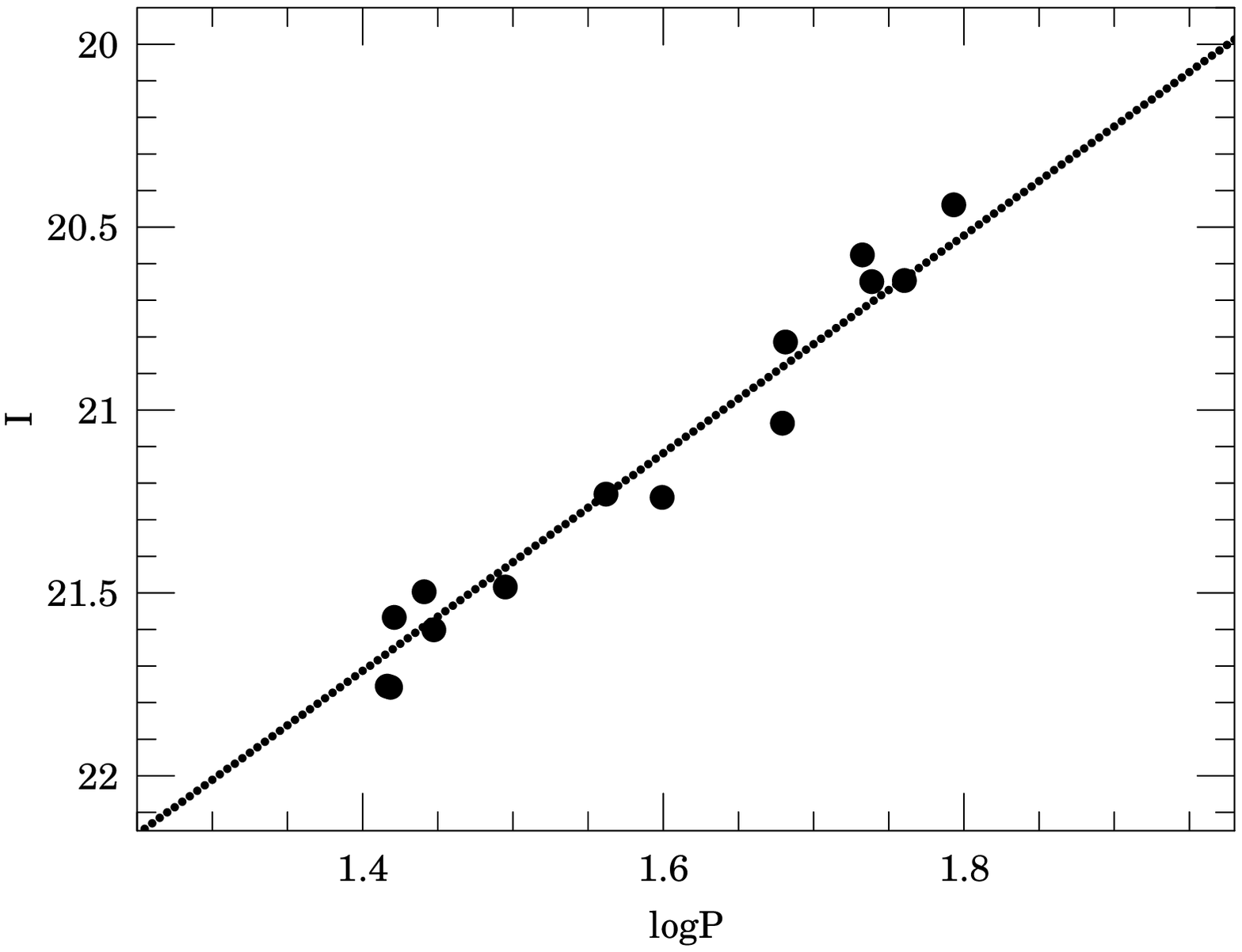}
\caption{Same as Fig. 3, for the I band.
}
\end{figure}

\begin{figure}[htb]
\vspace*{15cm}
\includegraphics{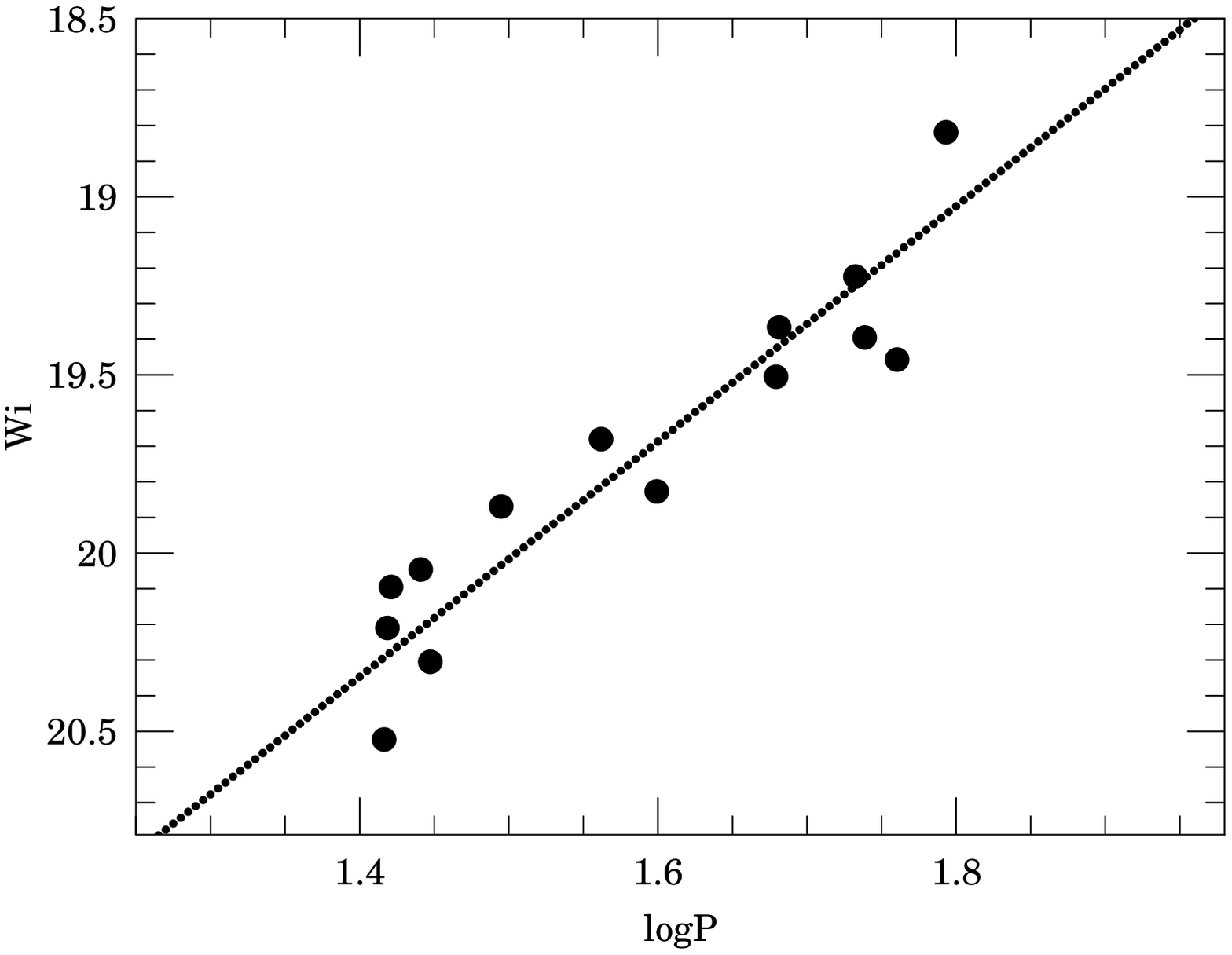}
\caption{Same as Fig. 3, for the reddening-independent (V-I) Wesenheit 
magnitudes. 
}
\end{figure}

\end{document}